\documentclass[11pt,twoside]{article}
\usepackage{asp2006}
\usepackage{epsf}
\usepackage{graphics}
\usepackage{lscape}
\newcommand{\be}{\begin{equation}}
\newcommand{\ee}{\end{equation}}
\newcommand{\bea}{\begin{eqnarray}}
\newcommand{\eea}{\end{eqnarray}}

\def\edcomment#1{\iffalse\marginpar{\raggedright\sl#1\/}\else\relax\fi}
\reversemarginpar
\begin{document}

\title{Cosmic ray transport in MHD turbulence}

\author{Huirong Yan \altaffilmark{1} and A. Lazarian \altaffilmark{2}}

\altaffiltext{1}{Canadian Institute of Theoretical Astrophysics, 60 St. George Street, Toronto, ON M5S 3H8, Canada; yanhr@cita.utoronto.ca}
\altaffiltext{2}{Astronomy Department, University of Wisconsin, Madison, WI 53706}
\begin{abstract}
Numerical simulations shed light onto earlier not trackable problem of magnetohydrodynamic (MHD) turbulence. They allowed to
test the predictions of different models and choose the correct ones.   Inevitably, this progress calls for revisions in the picture of cosmic ray (CR) transport.
It also shed light on the problems with the present day numerical modeling of CR. 
In this paper we focus on the analytical way of describing CR propagation and scattering, which should be used in synergy with the numerical
studies. In particular, we use recently established scaling laws for MHD modes
to obtain the transport properties for CRs.  We include nonlinear effects arising from
 large scale trapping, to remove the 90 degree divergence. We determine how the efficiency of the scattering and CR mean free path depend on the 
 characteristics of ionized media, e.g. plasma $\beta$, Coulomb collisional mean free path. Implications for particle transport in interstellar 
medium and solar corona are discussed. We also examine the perpendicular transport of CRs. Perpendicular transport depends on the comparison of parallel mean free path and the injection scale of the turbulence, as well as the Alfv\'enic Mach number. Normal turbulence does not allow subdiffusion unless there are slab waves. The critical scale below which subdiffusion applies is provided. These results can be used to compare with the numerical calculations, provided that these calculations use the structure of magnetic
field which is consistent with the numerical studies of MHD turbulence.
\end{abstract}

\section{Introduction}
The propagation and acceleration
of cosmic rays (CRs) is governed by their interactions
with magnetic fields. 
The perturbations of turbulent magnetic field can be accounted for by 
direct numerical scattering simulations (see Giacalone \& Jokipii 1999) or by
 quasi-linear theory, QLT (see Jokipii 1966, Schlickeiser 2002). The problem
with direct numerical simulations of scattering is that the present-day
MHD simulations have rather limited inertial range. As a result, they are
sufficient to test theoretical expectations (see Cho \& Lazarian 2003) of the power spectra and its anisotropy, but are
not adequate to study the particle scattering. Indeed, if one directly uses the results
of MHD simulations, it is likely that the scattering will be happening by magnetic
perturbations beyond the inertial range. 

Alternative way would be creating magnetic fields in synthetic data cubes. However, both theory (Goldreich \& Sridhar 1995, henceforth GS95) 
and numerical simulations (Cho \& Vishniac 2000, Maron \& Goldreich 2001, Cho, Lazarian \& Vishniac 2002, Muller \& Biskamp 2000, Cho \& Lazarian 2003
and ref. therein) show that the Alfv\'enic modes exhibit scale-dependent anisotropy, i.e. the degree of anisotropy of eddies changes with scale. 
The complication here is that this anisotropy can be seen only in the system of reference connected with the local magnetic field, which makes
the accepted procedures of generating random fields in Fourier space inapplicable. As CRs do sample local magnetic fields, it seems essential to use
the correct description of this field, which presents an exciting, but still unsolved problem for numerical modeling of CR propagation.
In this situation, analytical calculations are indispensable. These studies should also provide one with guidance in interpreting the present-day CR propagation simulations
and designing the new ones. 

So far the most frequently used analytical formalism is QLT. While QLT provides simple physical insights into scattering, it is known
to have problems. For instance, it fails in treating $90^\circ$ scattering and perpendicular transport. 
Indeed, many attempts have been made to improve the QLT and various non-linear
 theories have been attempted (see Dupree 1966, V\"olk 1973, Jones, Kaiser \& Birmingham 1973, 
Goldstein 1976, Matthaeus et al. 2003, Shalchi 2005). Most of the analysis so far are confined to 
slab model and 2D of MHD perturbations. We feel that it is important to extend the work on non-linear treatment of CR scattering to models of large scale 
MHD turbulence. Here we use models that are motivated by
the recent studies of MHD turbulence (GS95, see Cho, Lazarian, Vishniac 2003 for a review 
and references therein). We showed in Yan \& lazarian (2002, 2004, henceforth YL02, 04) that 
scattering of CRs is dominated by fast modes. We shall reexamine our earlier conclusions. More important, by solving the $90^\circ$ scattering, we shall determine how CR mean free path varies in different environments. 

We shall also address the problem of perpendicular transport of CR,
which is another important problem in which QLT encounters serious difficulties. 
Indeed, the idea of CR transport in the direction perpendicular to the mean
magnetic field being dominated by the field line random walk (FLRW, Jokipii 1966, 
Jokipii \& Parker 1969, Forman et al. 1974) only applies to a restricted situation 
where the turbulence perturbations are small and CRs' mean free paths
are larger than the injection scale of MHD turbulence. 
It was speculated that in some regime the particle motions can be subdiffusive (K\'ota \& Jokipii 2000, Mace et al 2000, Qin at al. 2002a, Shalchi 2005, Webb et al. 2006).
 This could indicate a substantial shift in the paradigm of CR transport.   
How realistic is the subdiffusion in the presence of turbulence injected on 
large scales? In this paper we shall use the tested model of turbulence to study the issue of subdiffusion. 
\section{Nonlinear transport in MHD turbulence}

The basic assumption of quasi-linear theory is that particles follow unperturbed orbits until they are scattered off by $\sim 90^\circ$. In reality, particle's pitch angle varies gradually with the variation of magnetic field due to conservation of adiabatic invariant. The average uncertainty of parallel speed $\Delta v_\parallel$ is $\Delta v_\parallel\simeq{v_\perp}\left(<\delta B_\parallel^2>/B_0^2\right)^{\frac{1}{4}}$, where $B_0$ is mean magnetic field strength. Perturbation $\delta B_\|$ exists owing to slow modes in incompressible turbulence and high $\beta\equiv \frac{P_{gas}}{P_{mag}}$ turbulence. In low $\beta$ compressible turbulence, $\delta B_\|$ arises from fast modes.  
This results in broadening of the resonance. Because of the dispersion of the pitch angle $\Delta\mu$ and therefore parallel speed $\Delta v_\|$, 
the guiding center is perturbed about the mean position $<z>=v\mu t$ as they move along the field lines.

\begin{figure}[!ht]
\plottwo{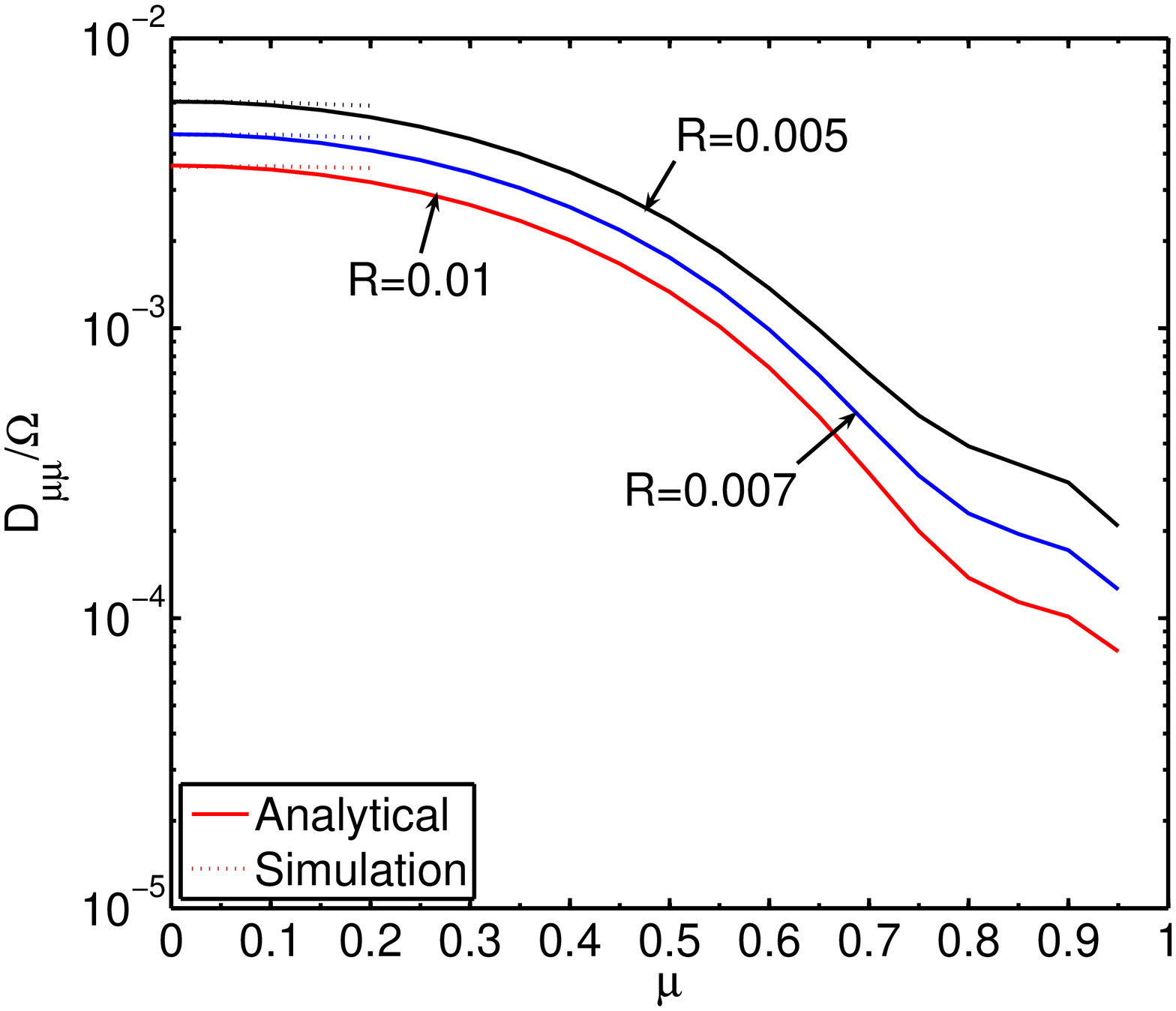}{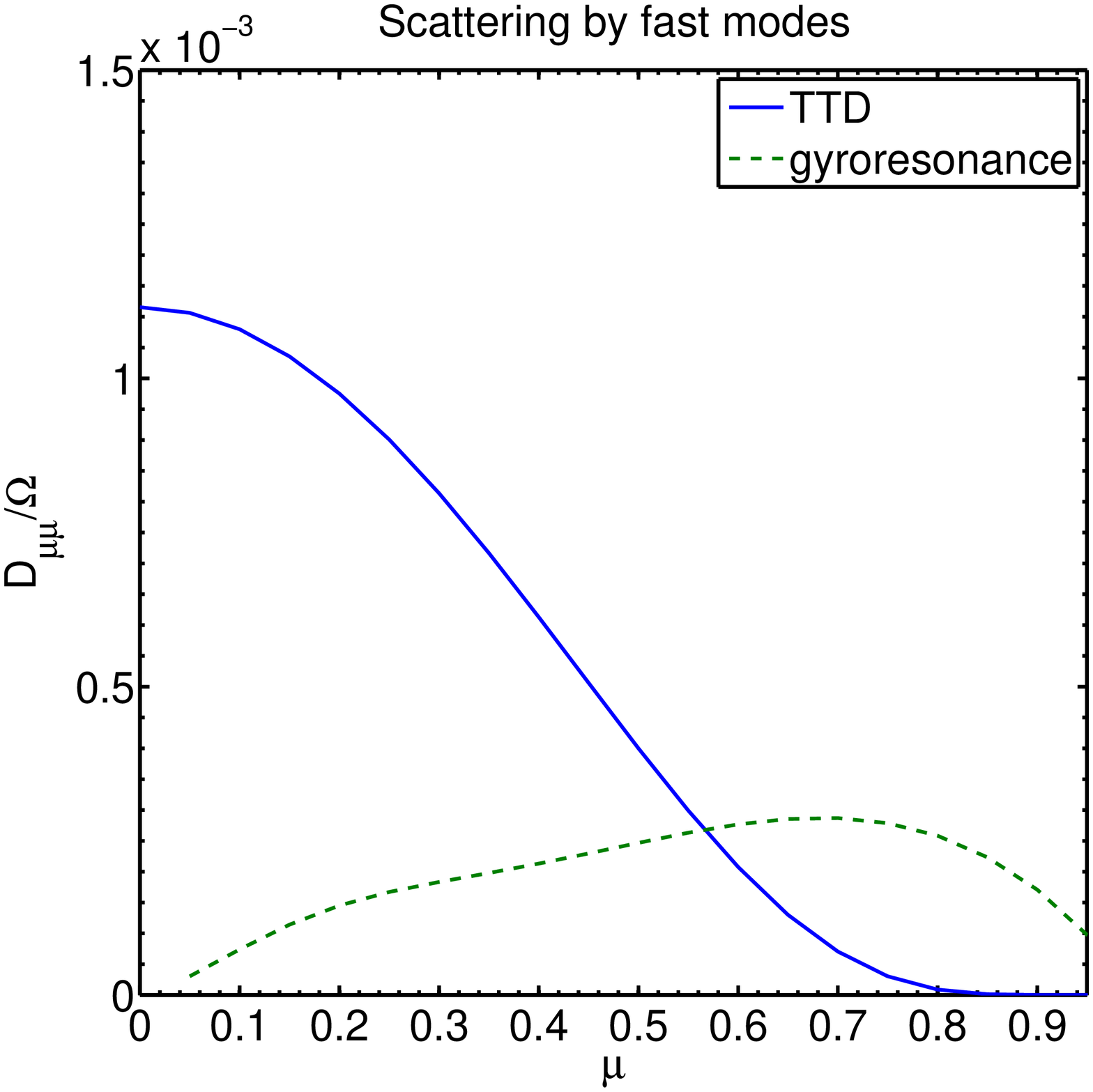}
\caption{Scattering of CRs in strong turbulence (Alfv\'enic Mach number $M_A=\delta B^2/B_0^2\simeq 1$). Left: scattering coefficient in strong incompressible turbulence; right: scattering coefficient in fast mode turbulence.}
\label{Dmumu}
\end{figure}
 The characteristic phase function $e^{ik_\|z(t)}$ deviates from that for plane waves, assuming the guiding center has a Gaussian distribution along the field line, we get the broadened resonance function
\be
R_n(k_{\parallel}v_{\parallel}-\omega\pm \Omega)=\frac{\sqrt{2\pi}}{2\pi|k_\|\Delta v_\||}\exp\left[-\frac{2(k_\|v \mu-\omega+n\Omega)^2}{k_\|^2\Delta v_\|^2}\right],
\ee
in contrast to the $\delta(k_{\parallel}v_{\parallel}-\omega\pm \Omega)$ function resonance in QLT. In the equation, $k_\|$ is the component of wave vector ${\bf k}$ projected along the local magnetic field, $\omega$ is the wave frequency and $\Omega$ is the Larmor frequency of the particle. The corresponding pitch angle diffusion can then be obtained by replacing the $\delta$ function with the new resonance function. For gyroresonance ($n=\pm 1,2,...$), the result is similar to that from QLT. This is reasonable as the gyroresonance works on a local scale so that the effect of large scale trapping is negligible.

On the other hand, the dispersion of the $v_\parallel$ means CRs with a much wider range of pitch angle can be scattered through TTD (transit time damping, n=0) by the compressible modes. In incompressible medium, TTD happens with pseudo-Alfv\'en modes (the incompressible limit of slow modes). The scattering by TTD produces a flat curve at large pitch angle (including $90^\circ$, see Fig.\ref{Dmumu}), which is confirmed by our preliminary result from simulations (Beresnyak et al. in preparation). 

Recent studies have shown that Alfv\'en (and slow) modes exhibit scale-dependent anisotropy and
follow GS95 
relations. The mixing motions associated with
Alfv\'enic turbulence induce the scale-dependent anisotropy $k_\|\sim k_\bot^{2/3}L^{-1/3}$ on slow modes,
which on their own would evolve on substantially longer time scale. Fast modes in low $\beta_p$ plasma, on the other hand, 
develop on their own, as their phase
velocity is only marginally affected by mixing motions induced
by Alfv\'en modes.   According
to Cho \& Lazarian (2002), fast modes 
follow an isotropic ``acoustic"
cascade with one dimensional energy spectrum $
W(k)\propto k^{-\frac{3}{2}}$. We adopt these models of MHD turbulence in the paper.

Although the $90^\circ$ divergence can be removed in the nonlinear treatment, the mean free path would be still unrealistically large if only Alfv\'en and slow modes are counted. This is because at small pitch angle (propagating nearly parallel to the magnetic field), only gyroresonance applies, which is dominated by the interaction with fast modes (YL02, 04). 

In reality, compressible turbulence is subjected to damping. The damping depends on the pitch angle between the {\bf k} vector and {\bf B} field (see Fig.\ref{mfpath}{\it left}). Accordingly, for CR with Larmor radius $r_L\ll L$, the injection scale of the turbulence, we obtain the pitch angle diffusion coefficient (Yan \& Lazarian 2007, henceforth YL07): 
\bea
D_{\mu\mu}=\frac{\sqrt{2\pi}}{8}v(1-\mu^2)^{1.5}\exp\left[-\frac{2\mu^2}{\Delta \mu^2}\right]\int_0^1 d\eta \sqrt{k_{max}(\eta)/L}\eta(1-\eta^2)
\eea
where $k_{max}(\eta)$ is the cut-off wave number at $\eta$, $\eta=\frac{k_\|}{k}$ is the cosine of wave pitch angle. Given $D_{\mu\mu}$, the mean free path can be obtained by $l_{mfp}=3/4\int_0^1d\mu \frac{v(1-\mu^2)^2}{D_{\mu\mu}}$.
\section{Implications for Galactic CR transport and solar flare}
\begin{figure}
\plottwo{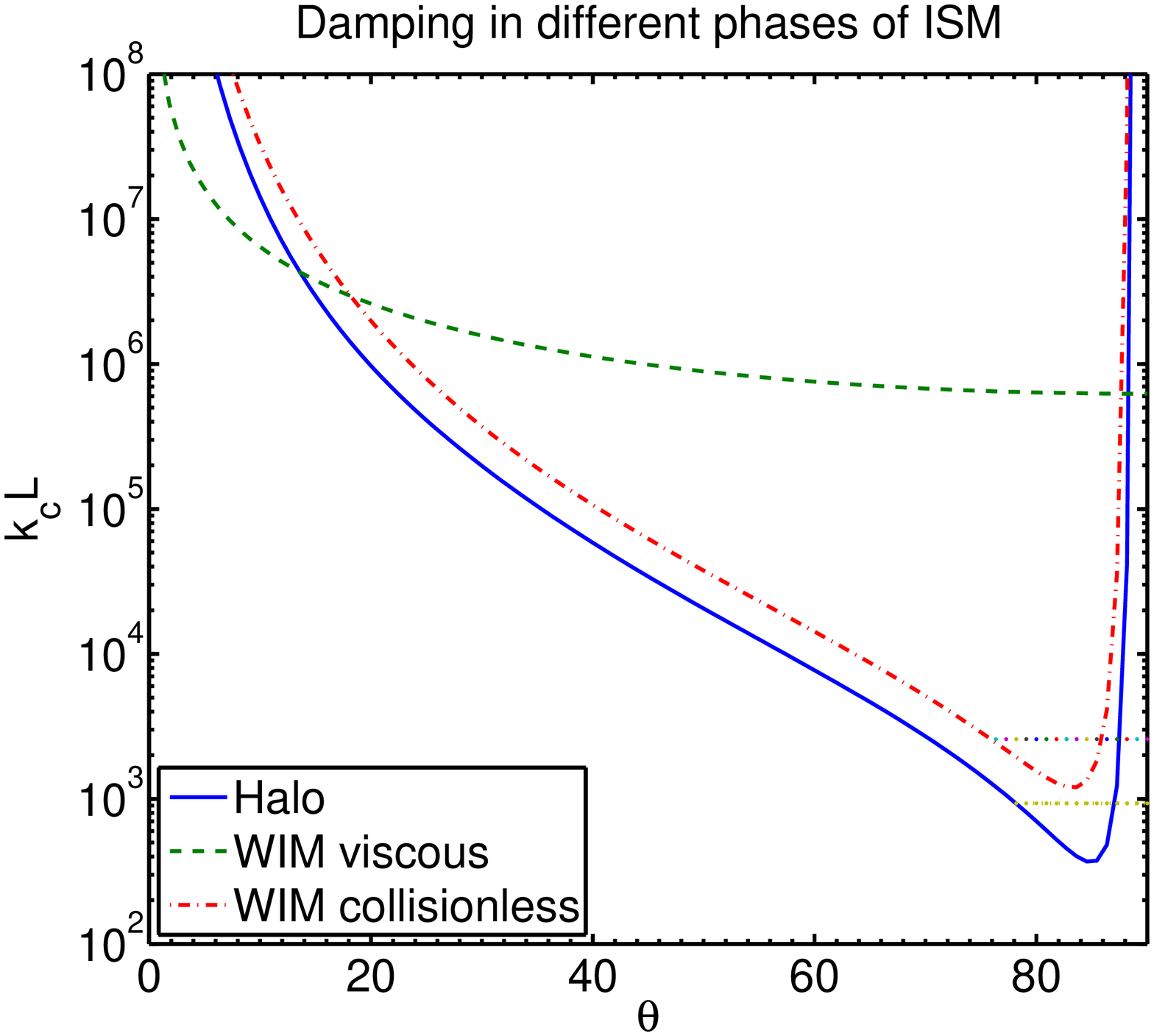}{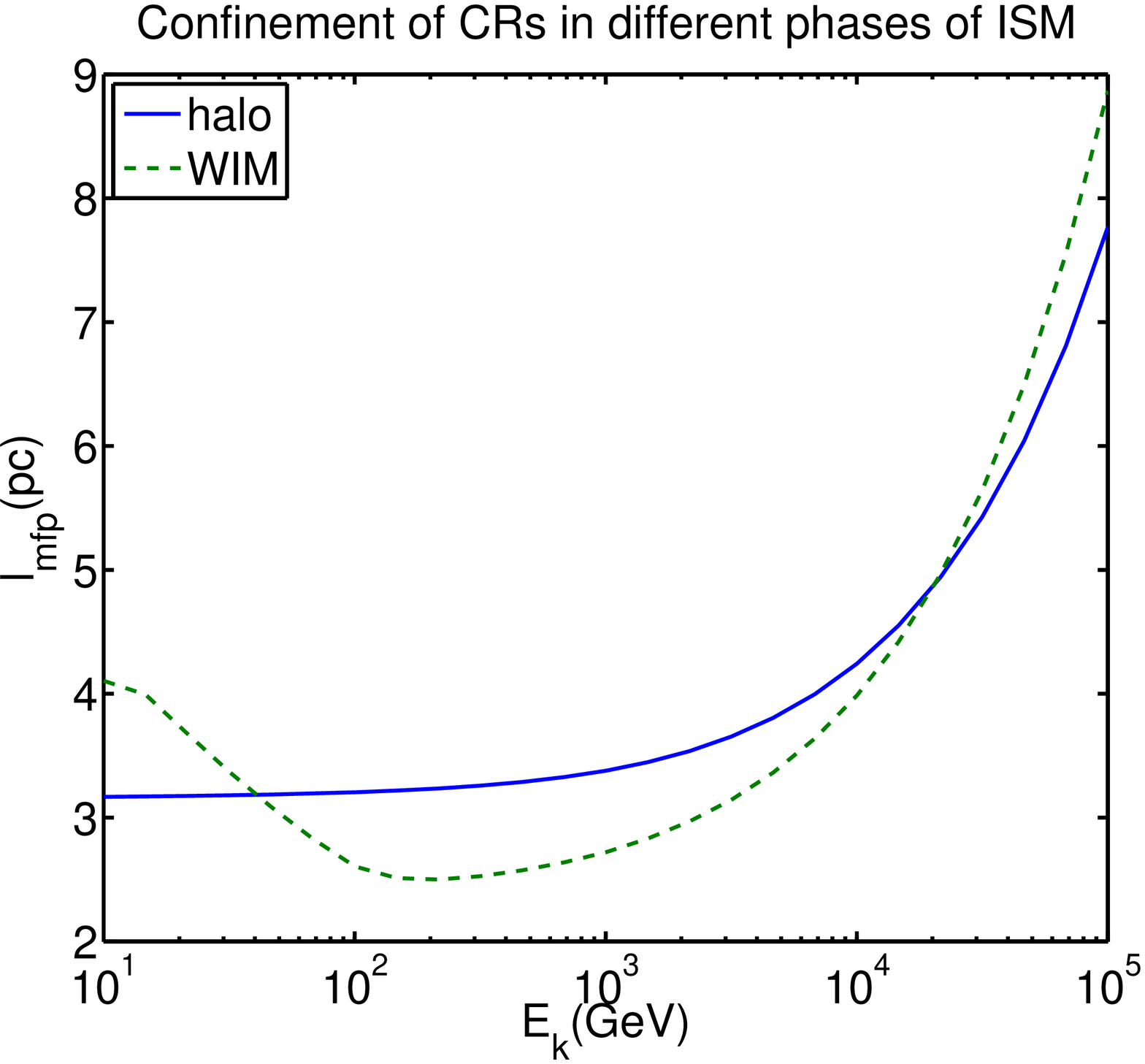}
\caption{{\it Left}: The turbulence truncation scales in Galactic halo and warm ionized medium (WIM). The damping curve becomes flat around $90^\circ$ in halo due to field line wandering; For WIM, both viscous and collisionless damping are applicable. {\it right}: The mean free paths in halo (solid line) and WIM (dashed line).}
\label{mfpath}
\end{figure}
The scattering by fast modes is influenced by the medium properties as fast modes are usually subjected to damping, which varies from place to place. 
YL04 has shown that the scattering properties for different ISM phases are different from each other. With the scattering at $90^\circ$ known, 
we can make quantitative predictions for the mean free path of CRs in different phases. 

Galactic halo is in collisionless regime. The damping curve is given in Fig.\ref{mfpath}{\it left}. As we see, the damping decreases with the wave pitch angle $\theta$. The mean free paths for moderate energy CRs are almost constant. This is due to the fact that gyroresonance changes
 marginally with the CR energy (see YL07 for details). For higher energy CRs, the influence of damping saturates, and mean free path begins increasing with energy.
In warm ionized medium (WIM) both viscous damping and collisionless damping are important (Fig.\ref{mfpath}{\it left}). We see that at low energies ($\leq 100$GeV), the CR's mean free path decreases with the energy. This is because of the influence of viscous damping on gyroresonance.

\begin{figure}
\plottwo{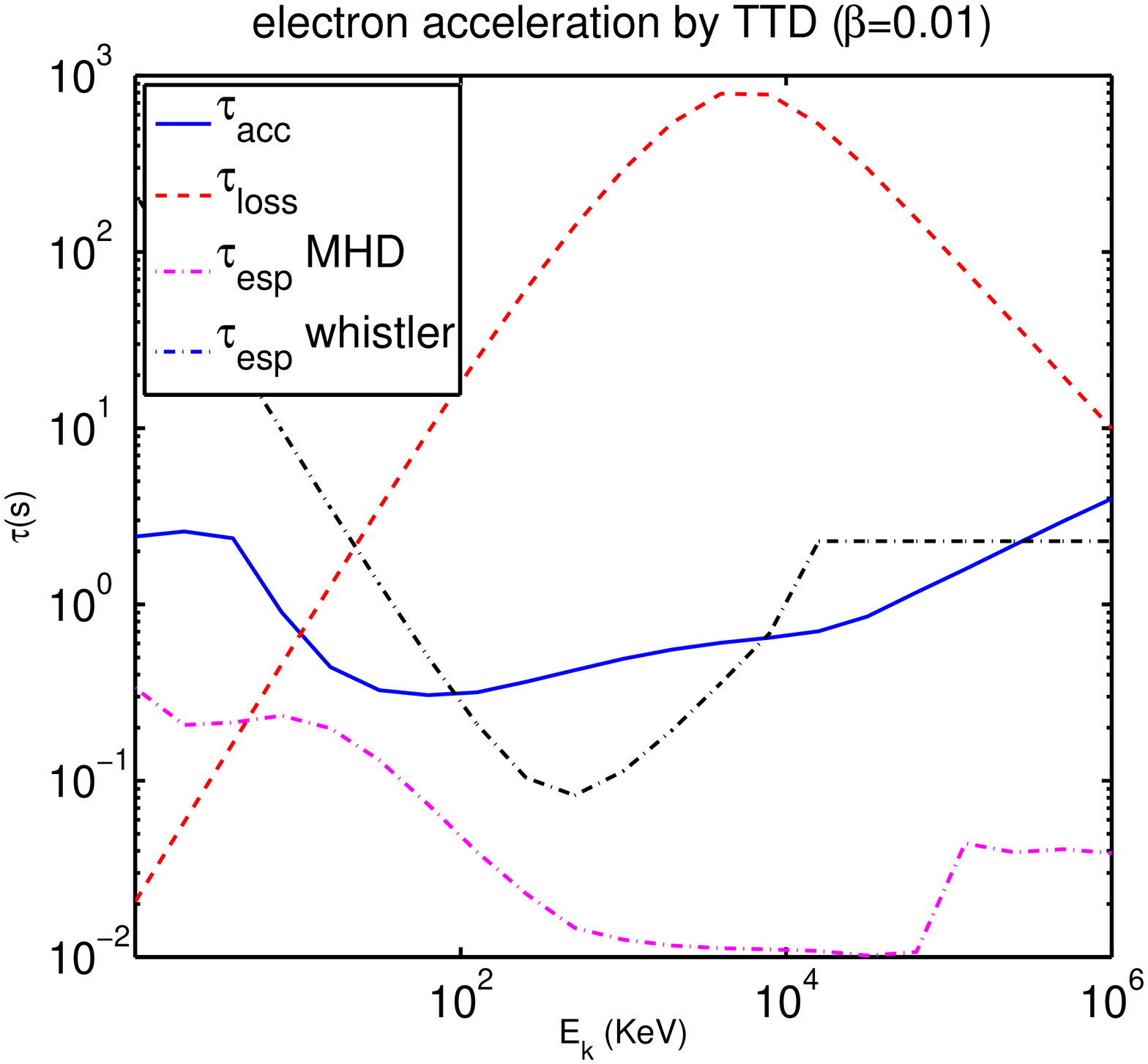}{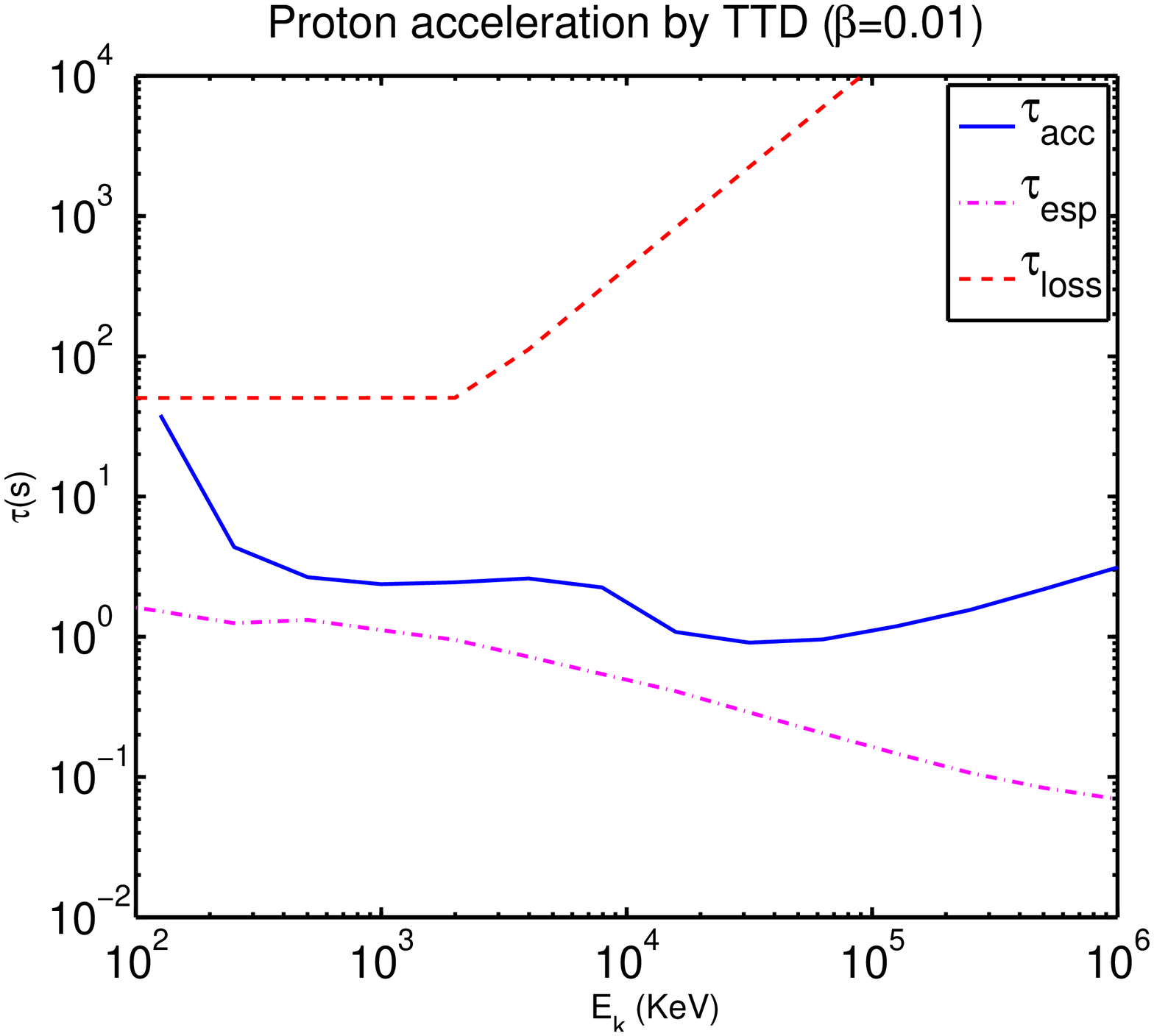}
\caption{Timescales involved in the acceleration in solar corona. The solid lines refer to the 
acceleration timescale. The dashed lines represent energy loss time scale. The dashdot lines give 
the escaping timescales.}
\label{solar}
\end{figure}

Since stochastic acceleration is intimately related to the scattering process, we applied our result to acceleration for solar flares (Yan, Lazarian \& Petrosian 2007). We found that fast modes can provide efficient acceleration for both protons and electrons (see Fig.\ref{solar}). In the case of protons, confinement of the particles can be achieved by fast modes as well, which is essential for the acceleration. 

\section{Perpendicular transport}
In the three-dimensional turbulence, field lines are diverging away due to shearing by Alfv\'en modes. Since the Larmor radii of CRs are much larger than the minimum scale of eddies $l_{\bot, min}$, CRs will not be confined to the same field line. The cross-field transport thus results from the deviations of field lines as well as field line random walk on large scales ($>{\rm min}[L/M^3_A,L]$).

{\it High $M_A$ turbulence}: In high $M_A$ turbulence, magnetic field becomes dynamically important on the scale $l_A=L/M_A^3$ (Lazarian 2006). If $\lambda_\|\ll l_A$, the stiffness of B field is negligible so that $D_\bot=D_\|$. If $\lambda_\|\gg l_A$, the diffusion is controlled by the straightness of the field lines, and $D_\bot=1/3l_Av$. 

{\it Low $M_A$ turbulence}: In this situation the steps in perpendicular 
direction are of $L M_A^2$ length. To diffuse over a distance R with random walk of
$LM_A^2$ one requires $(R/LM_A^2)^2$ steps. Therefore the perpendicular diffusion coefficient 
is $R^2/(R^2/D_{\parallel}M_A^4)=D_{\|}M_A^4$. In the case $\lambda_\|>L$, the time of the individual step is $L/v_\|$, then $D_\perp=1/3Lv M_A^4$. 
This is similar to the case discussed in the FLRW model (Jokipii 1966). However, we obtain the dependence of $M_A^4$ instead of their $M_A^2$ scaling. 

Subdiffusion happens when particles are restricted to the magnetic field lines and the perpendicular transport 
is solely due to the random walk of field lines (see K\'ota \& Jokipii 2000).  We do not have it in our earlier 
models. However, we may have sufficiently strong slab mode and it can cause 
subdiffusion. It is a very restrictive case. 
Consider a slab component. This component causes random walk of field 
lines in the plane perpendicular to the mean magnetic field direction. 
For a power-law spectrum $E(k)\propto (kl_{slab})^{-\alpha}$, the critical scale up to which this component may 
induce random walk in the presence of Alfv\'enic field wandering is
$z_{crit}=(kl_{slab})^{-\alpha/2}\sqrt{Ll_{slab}}M_A^{-2}A_s$,
where $A_s=\delta B_s/B_0$ is the dimensionless amplitude of the slab modes on scale $l_{slab}$. The dominant wavelength can be determined by maximizing 
$\delta b_k k^{-1}$. Naturally, the constrain for this is that
$k^{-1}<z_{crit}$. Otherwise no random walk has a meaning. Combining it with the expression of $z_{crit}$, we get $\alpha\leq \alpha_{crit}=2$. Detailed derivation provides $D_\bot=A_sl_{slab}D_\|^{\frac{1}{2}}$. 
\section{Summary}
The cosmic ray transport is modified in 
accordance with the progress of understanding of MHD turbulence. 
Implementing the tested model of turbulence, we found that fast modes 
dominate CR scattering if the turbulence energy is injected on large scales. On the other hand, the insights we gained through analytical studies give us guidance for designing future numerical experiments to test CR transport theories. 

Damping of fast modes leads to more dependencies of CR transport on the medium properties, different from the scattering by Alfv\'en modes.    
Transient time damping (TTD) provides an important means of cosmic ray transport. Unlike gyroresonance, nonlinear broadening should be accounted. The $90^\circ$ degree scattering can be realized by the broadened TTD, which ensures finite mean free paths for CRs. Gyroresonance with 
fast modes prevent streaming of moderate to high energy CRs, which 
cannot be confined by streaming instability.

Perpendicular transport is diffusive, controlled by the diverging of field lines and field line random walk. The perpendicular diffusion depends on the CR's parallel mean free path and the strength of magnetic perturbation. Subdiffusion only applies below some critical scale if slab waves exist. 

In terms of numerical simulations of CR propagation, our findings of the dominance of fast modes, which, according to numerical studies are much more isotropic, compared to
Alfv\'en modes, may suggest that the present-day codes can capture important aspects of CR physics, provided that the damping of fast modes is properly taken into account.

\end{document}